\documentclass[aps,pre,12pt]{revtex4-1}

\newcommand\numberthis{\addtocounter{equation}{1}\tag{\theequation}}

\usepackage{amsmath}
\usepackage{amssymb}
\usepackage{graphicx}
\usepackage{color}
\begin{document}
\title{Emergent user behavior on Twitter modelled by a stochastic differential equation}
\author{Anders Mollgaard and Joachim Mathiesen}
\affiliation{Niels Bohr Institute, University of Copenhagen, Blegdamsvej 17, DK-2100 Copenhagen, Denmark}
\begin{abstract}
Data from the social-media site, Twitter, is used to study the fluctuations in tweet rates of brand names. The tweet rates are the result of a strongly correlated user behavior, which leads to bursty collective dynamics with a characteristic 1/f noise. Here we use the aggregated "user interest" in a brand name to model collective human dynamics by a stochastic differential equation with multiplicative noise. The model is supported by a detailed analysis of the tweet rate fluctuations and it reproduces both the exact bursty dynamics found in the data and the 1/f noise.
\end{abstract}

\maketitle
\section{Introduction}
In the online era, humans are connected in real time on global scales. Local or seemingly local information is instantaneously shared across geographical boundaries. In particular, social online media  have become an important platform for the sharing of information and have allowed for detailed studies of the coherent behavior of humans on a global scale \cite{king2011ensuring, osborne2014facebook, hermida2014sourcing, mathiesen2013excitable,alstott2014powerlaw,garas2012emotional}. The popular microblogging platform Twitter is a good source for such studies for two reasons. First, Twitter is more about providing news updates than developing social networks \cite{kwak2010twitter, myers2014information}. User behavior is therefore to a large extent influenced by information available via other information channels in society. Secondly, users respond to available information by submitting short public messages, "tweets", of up to 140 characters that may be seen as proxies for the public interest. Recent research on Twitter has used the activity levels in forecasting real-world events including fluctuations of stock market prices \cite{Sakaki:2010:EST:1772690.1772777}, real-time detection of the location and spread of earthquakes hitting populated areas \cite{DBLP:journals/corr/abs-1010-3003}, and for sentiment analysis and opinion mining \cite{DBLP:journals/corr/abs-1003-5699}.

In a recent paper \cite{mathiesen2013excitable}, fluctuations in the tweet rates of 92 brand names are shown to be distributed with a power law tail with an exponent of $-2.9\pm0.4$(SD). The broad tail of the distribution is characteristic for bursty activity levels. It is moreover found that the power spectral density of the tweet rate signals are described by a power law with an exponent of $-1.0\pm0.4$(SD). This so called, "1/f noise", is found in a range of complex systems including heartbeats (\cite{kobayashi19821}), DNA base sequences (\cite{voss1992evolution}) and condensed matter systems (\cite{weissman19881}), and it is interpreted as a sign of a pronounced memory in the systems (\cite{keshner19821}). We attribute the power spectral density and the broad distribution in the tweet rate fluctuations to a strong correlation on a global scale in the collective human dynamics.

In this paper we consider the global user interest in a brand, which in our definition is the likelihood for a tweet to mention the brand name. The global interest in a topic is expected to change in a continuous and random fashion as the result of many independent events in society.
We shall therefore describe the global user interest by a stochastic differential equation (SDE), which we derive by analyzing the fluctuations in the tweet rate. The SDE predicts simultaneously the power law exponents of the tweet rate distribution and the strong memory in the temporal variation of the tweet rates.

The paper is organized as follows, first we briefly describe the data acquired from Twitter and explain how the data is turned into a tweet rate. Then we introduce a method for analyzing the fluctuations in the the tweet rate and demonstrate how it works on a generic signal. Our method is applied to data and supports an SDE with multiplicative noise. Finally, we show that the noise term in the SDE reproduces the power law distribution of the tweet rate as well as the power spectral density of the temporal signal.

\begin{figure}[htb]
	\centering
	\includegraphics[width=0.4\textwidth]{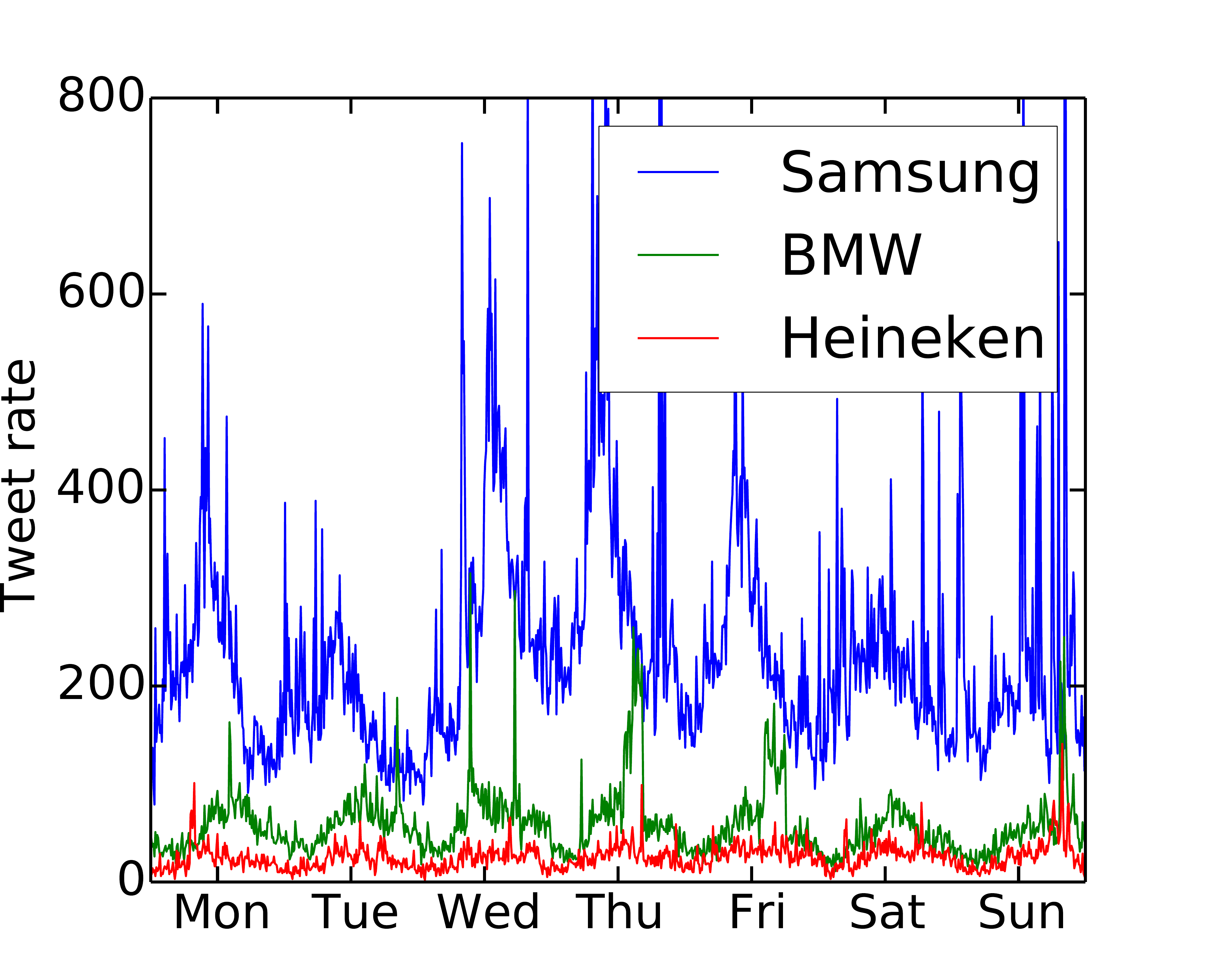}
	\caption{\textbf{Tweet rate signals} We show the number of tweets measured in a time window, $\Delta T=10\mathrm{min}$, for a few brands. Note the regular daily variation and the irregular bursty behavior.}
	\label{fig:data characteristics}
\end{figure}

\section{Methods}
\subsection{Data collection and time signals}

We used the public REST API by Twitter to collect tweets containing one of seven brand names during a time period in the fall 2012 and the spring 2014 (see Supporting Information sections S1 and S2 for data and a description of time periods). The brand names considered are "Samsung", "Pepsi", "Heineken", "Gucci", "Starbucks", "BMW", and "Google". In the analysis, we chose to use international brand names for a number of reasons. First, the brand names are used globally and the users posting tweets about the brands in general transcend local communities. Secondly, the brands are sufficiently popular that a continuous and robust stream of tweets exists.

From the tweets collected, we save the time $t_i$ where a tweet is posted. The index $i$ refers to the identification number of a given tweet. From the individual tweets, we form a time signal by summing over all tweets mentioning a given brand, $$s(t)=\sum_i \delta(t-t_i),$$ where $\delta(t-t_i)$ is the Dirac delta function. The time signal is turned into a tweet rate, $x(t)$, by dividing the time axis into windows of length, $\Delta T$, and summing the events in each window,
\begin{equation}\label{observed}
x(t) = \int_{t-\Delta T/2}^{t+\Delta T/2} s(t') dt'.
\end{equation}
A plot of tweet rate signals is shown for a few brands in Fig. \ref{fig:data characteristics}, where both a regular daily variation and an irregular bursty behavior on top are distinctly visible. Burstiness is known to be inherent to individual human dynamics \cite{Barabasi:2005} and to have an impact on information spreading \cite{horvath2014spreading}. Here we see that bursts also appear in the aggregated interest level of many users in a large-scale social organization.

\begin{figure}[htb]
	\centering
	\includegraphics[width=0.5\textwidth]{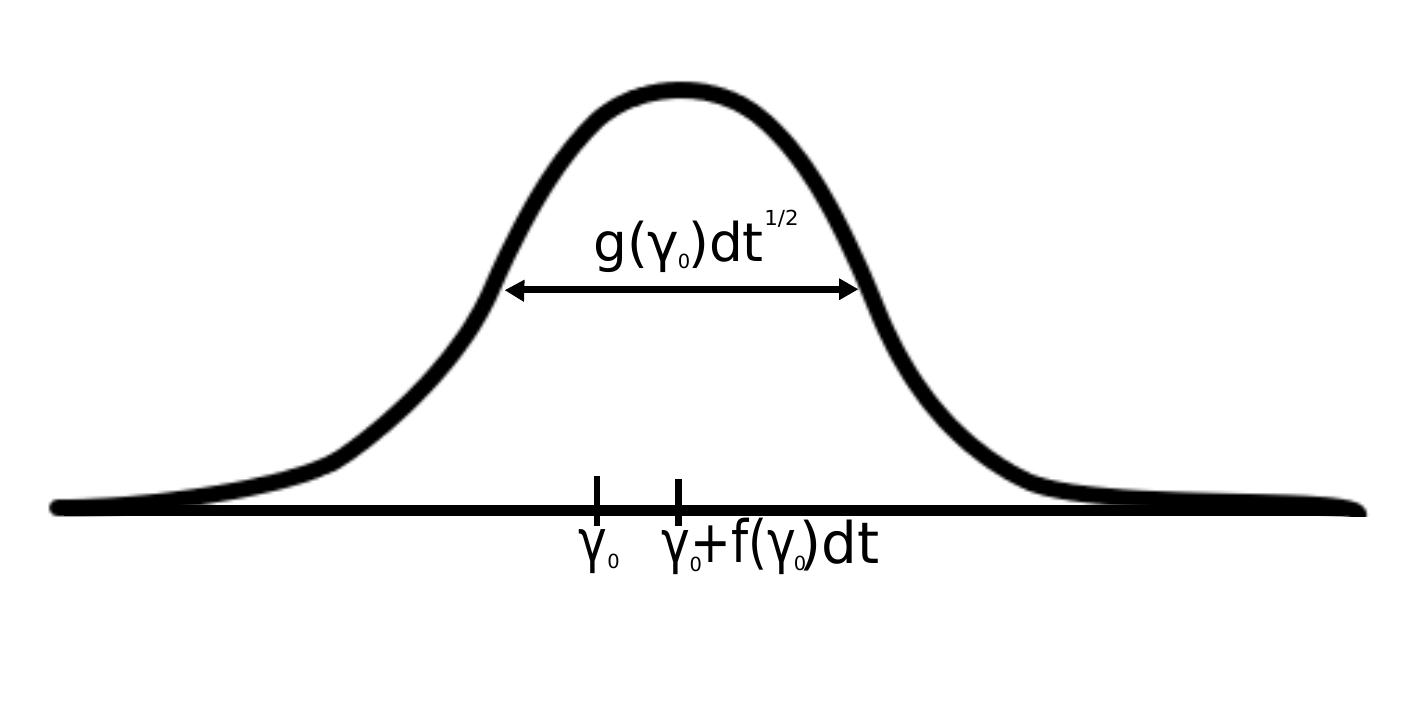}
	\caption{\textbf{Interpretation of the update formula in Eq. (\ref{eq:general sde}).} If the signal at some point takes the value $\gamma_0$, then a small time step, $dt$, later, its value will be realized from a Gaussian with a mean determined by $f(\gamma_0)$ and a spread determined by $g(\gamma_0)$. By performing statistics over many such realizations we may therefore obtain the drift and diffusion.}\label{fig:SDE interpretation}
\end{figure}

\subsection{The algorithm}
We now introduce a method to uncover the underlying stochastic properties of a time signal \cite{siegert1998analysis}. The method is based on the assumption that a signal, $\gamma(t)$, is generated by a stochastic differential equation (SDE) on the form

\begin{equation}\label{eq:general sde}
d\gamma = f\left(\gamma\right)dt + g(\gamma)dW.
\end{equation}

Here $dW$ is a random Gaussian variable with mean, $\left<dW\right>=0$, and variance, $\left<dW^2\right>=dt$. The first term in Eq. (\ref{eq:general sde}) gives the deterministic drift, while the second term gives the random diffusion. The differential equation will here be handled using Ito calculus. Note that the above equation is assumed to describe the dynamics of the global user interest and not the observed tweet rates given in Eq.~(\ref{observed}). Below we shall relate the two quantities.

It may be shown in the Fokker-Planck formalism \cite{gardiner2009stochastic} that if the variable takes the value, $\gamma_0$, at time, $t$, then at some small time step, $dt$, later, it will be a random variable from a Gaussian distribution with mean, $\gamma_0+f(\gamma_0)dt$, and spread, $g(\gamma_0)dt^{1/2}$ (see Fig.~\ref{fig:SDE interpretation}). It is therefore possible to get an estimate of $f(\gamma_0)$ and $g(\gamma_0)$ by binning all the signal values close to $\gamma_0$ and then construct the corresponding distribution of signal values one time step later. From this distribution one may read off the mean and the spread to get the estimates of $f\left(\gamma_0\right)$ and $g\left(\gamma_0\right)$. The procedure is then repeated over the whole range of realized signal values in order to estimate the functional forms of $f(\gamma)$ and $g(\gamma)$, respectively.

\begin{figure}[htb]
	\centering
	\includegraphics[width=0.35\textwidth]{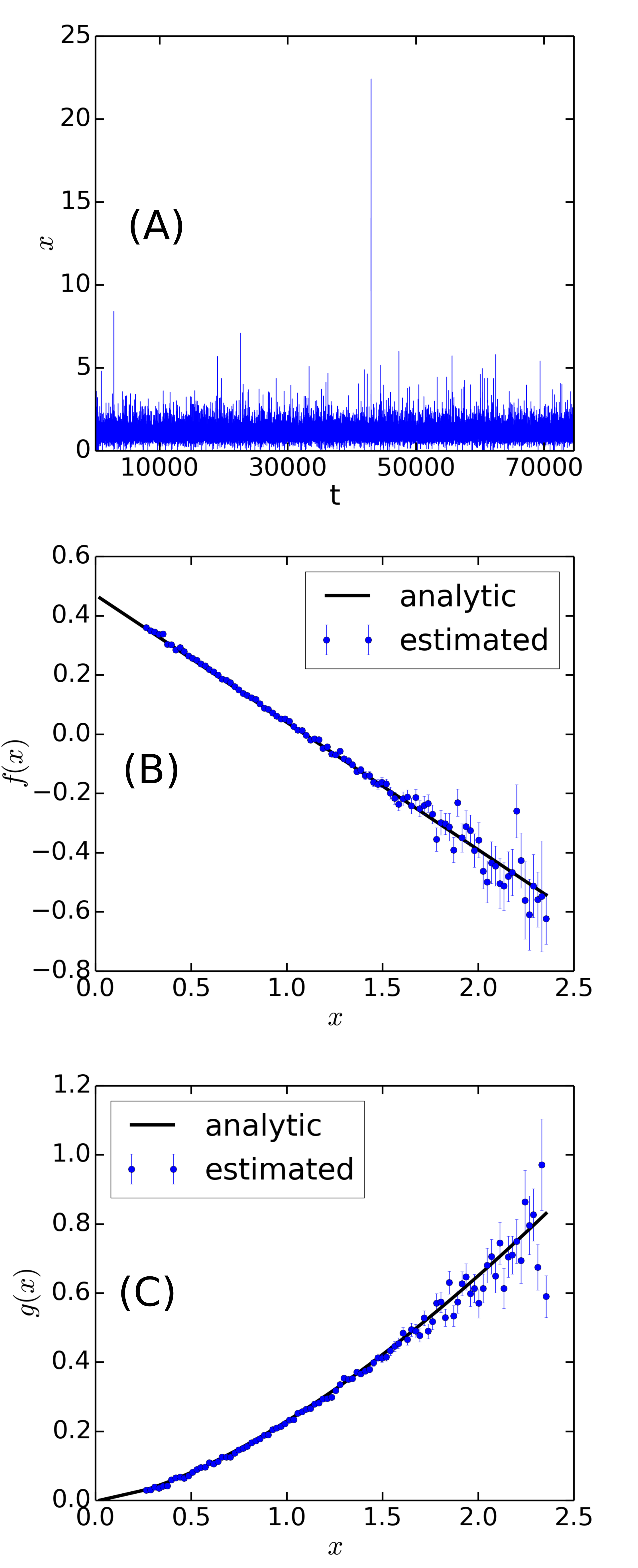}
	\caption{\textbf{Application of the algorithm to a generic signal.} A time series generated according to Eq. (\ref{eq:generic sde}) is shown in (A) and below we show the analytic drift, (B), and diffusion, (C), along with the functional forms estimated by the algorithm. }\label{fig:generic signal}
\end{figure}

In Fig. \ref{fig:generic signal} we show the result of applying the analysis to a signal generated by

\begin{equation}\label{eq:generic sde}
d\gamma = \left(0.47-0.43\gamma\right)dt+0.23\gamma^{1.5}dW,
\end{equation}
Comparing the analytical functions with the estimates, we get $R^2$-values of 0.98 and 0.97 for the drift and diffusion respectively. The functional form of the drift and diffusion used here are equivalent to the ones fitted for "Samsung" below, and to make the comparison complete, we have also used the same signal length, $N=74,646$, and time step, $dt=1$. Note that bins with less than 20 data points have not been included, due to the otherwise poor statistics, and therefore $\gamma$ only assumes values between 0.26 and 2.35. 

\section{Results}
\subsection{Model}
The stochastic differential equation, Eq.~(\ref{eq:general sde}) is formulated in the probability, $\gamma(t)$, for a random tweet to mention a specific brand and not in the tweet rate. We call  $\gamma(t)$ for the "global user interest". 
In fact, the expected number of tweets on a given topic, $\left<x(t)\right>_P$, in a time window, $\Delta t$, is given by the full number of tweets posted on Twitter within this time window, $A(t)$, times the probability for any such tweet to mention the given topic $\gamma(t)$,

\begin{equation}\label{average tweet rate}
\left< x(t) \right>_P = \gamma(t)A(t).
\end{equation}
Here the expectation value, $\left<\cdot\right>_P$, refers to the Poisson weighted average of all the possible realizations of the tweet rate. The actual tweet rate signal is one such realization drawn from a Poisson distribution

\begin{equation}\label{eq:signal interpretation}
x(t) =  \mathrm{Pois}\left(\gamma(t)A(t)\right). 
\end{equation}
The above equation summarizes the basic structure of our model: the observed signal, $x(t)$, is realized from a Poissonian with a mean given by the product between the user interest, $\gamma(t)$, and the activity, $A(t)$.

Within the activity, $A(t)$, we also include any factors depending on regional differences, since the global composition of active users is changing during a daily cycle. We will assume that $A(t)$ may be approximated as a deterministic and periodic function of time and that $\gamma(t)$ is reasonably described by the SDE in Eq. (\ref{eq:general sde}). The goal of the following data analysis is to find the functional form of the drift, $f(\gamma)$, and diffusion, $g(\gamma)$, of the SDE.

\begin{figure}[h!]
	\centering
	\includegraphics[width=0.5\textwidth]{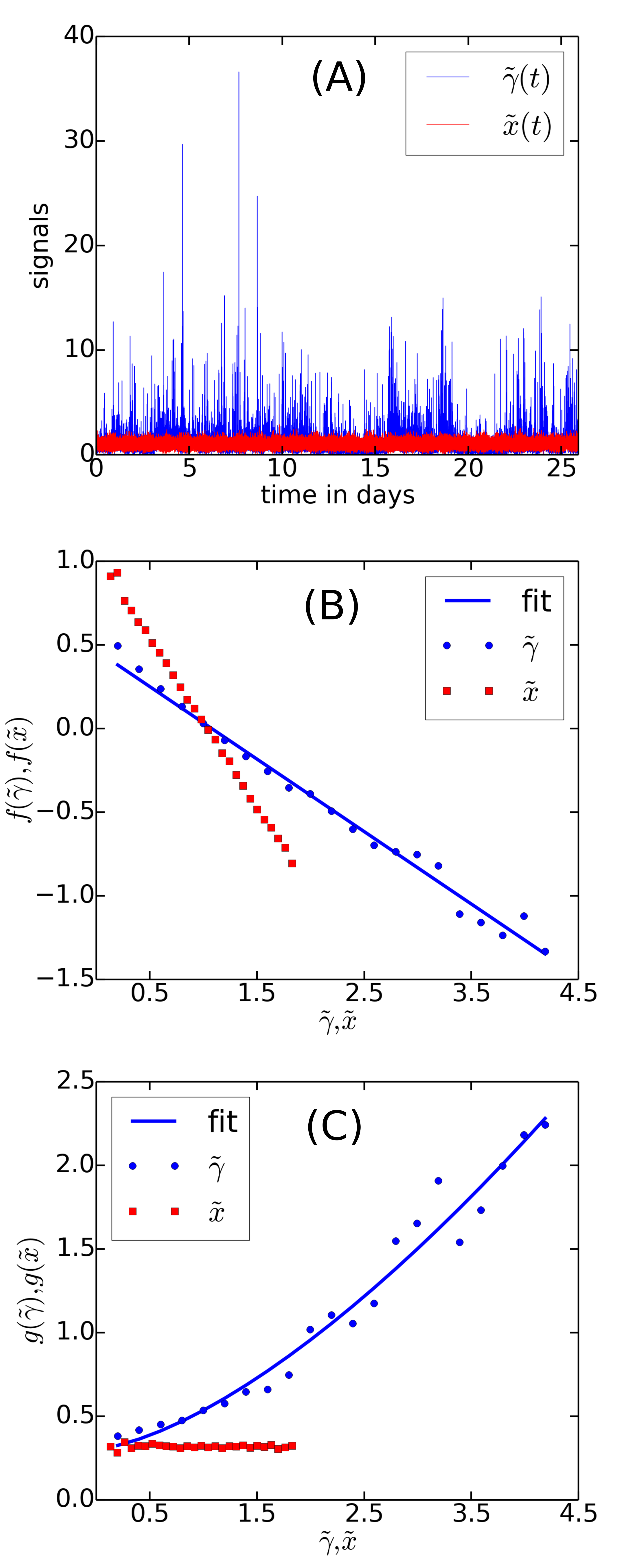}
	\caption{\textbf{Application of the algorithm to $\tilde{\gamma}(t)$ and $\tilde{x}(t)$ for the brand name "Samsung".} The two signals are shown in A and below we see the drift terms, B, and diffusion terms, C, estimated by the algorithm. Also shown are the fits of the functions in Eq. (\ref{eq:fit1}) and (\ref{eq:fit2}) to the estimated drift and diffusion of $\tilde{\gamma}(t)$.  }
	\label{fig:samsung analysis}
\end{figure}

\subsection{Data analysis}
It is problematic to apply the algorithm introduced in the methods section to the tweet rate signals, $x(t)$, since we only expect it to apply to the underlying tweet probability, $\gamma(t)$. We may reduce noise from the Poisson statistics by increasing the time window, $\Delta t$, thereby increasing the expected number of tweets and reducing the relative size of the Poisson noise. However, if we increase the time window too much, then we enter the domain of the mean field theory, where the time resolution is too low to see the dynamics of the $\gamma(t)$-fluctuations since they for larger times are dampened by the drift term. Also, for a limited time series we do not want to lower the time resolution, since it reduces the number of data points available for the analysis. In the following, we have chosen a time window of 30 seconds giving us approximately 80,000 data points for each brand. Unfortunately, this window size does not allow us to ignore the Poisson noise for any of the brands. We do however expect the time window to be small enough to resolve the important dynamics of the user interest.

The second problem that we face by applying the algorithm is the presence of the activity, $A(t)$, relating the observed signal, $x(t)$, to the signal of interest, $\gamma(t)$. In the following analysis we will assume that the activity is a deterministic function of time with a daily period. One would naturally expect it to also have a weekly variation along with a variation on slower time scales, but here we will be interested in time scales below the resolution of a day, why it makes sense to approximate the activity by a daily period.

We may estimate the daily variation by averaging $x(t)$ over many days to obtain a variable that is proportional to the activity

\begin{align*}
\left\langle x \right\rangle_D\!(t) &= \left\langle \mathrm{Pois}\left(\gamma(t)A(t)\right) \right\rangle_D, \\
&= \left\langle \gamma(t)A(t) \right\rangle_D, \\
&= \left\langle \gamma(t) \right\rangle_D \left\langle A(t) \right\rangle_D , \\
&= \left\langle \gamma \right\rangle_T A(t). \numberthis \label{define x_D}
\end{align*}
Here we introduced two more expectation values: $\left\langle \cdot \right\rangle_D$ is the average of repeated measurements at the same time of day and $\left\langle \cdot \right\rangle_T$ is the general time average. We have used that Poissonians sum to a Poissonian with an expectation value that is the sum of the individual expectation values. We have also used that $A(t)$ and $\gamma(t)$ are uncorrelated in our model, that $\gamma(t)$ is independent of absolute time and that $A(t)$ is a periodic function. In practice, the average is performed over 20 to 62 days of measurements and by smoothening data to a time resolution of 15 minutes. Using the obtained information, we may construct the variable 

\begin{align*}
\tilde{\gamma}(t) &\equiv \frac{x(t)}{\left\langle x \right\rangle_D\!(t)},\\
&= \frac{\mathrm{Pois}\left(\gamma(t)A(t)\right)}{ \left\langle \gamma \right\rangle_T A(t)  }, \numberthis \label{eq:tilde_gamma}
\end{align*}
which is proportional to $\gamma(t)$ if one averages out the Poisson noise

\begin{equation}
\left< \tilde{\gamma}(t) \right>_{P} = \frac{\gamma(t)}{\left<\gamma\right>_T}.
\end{equation}

\begin{figure}[htb]
	\centering
	\includegraphics[width=0.5\textwidth]{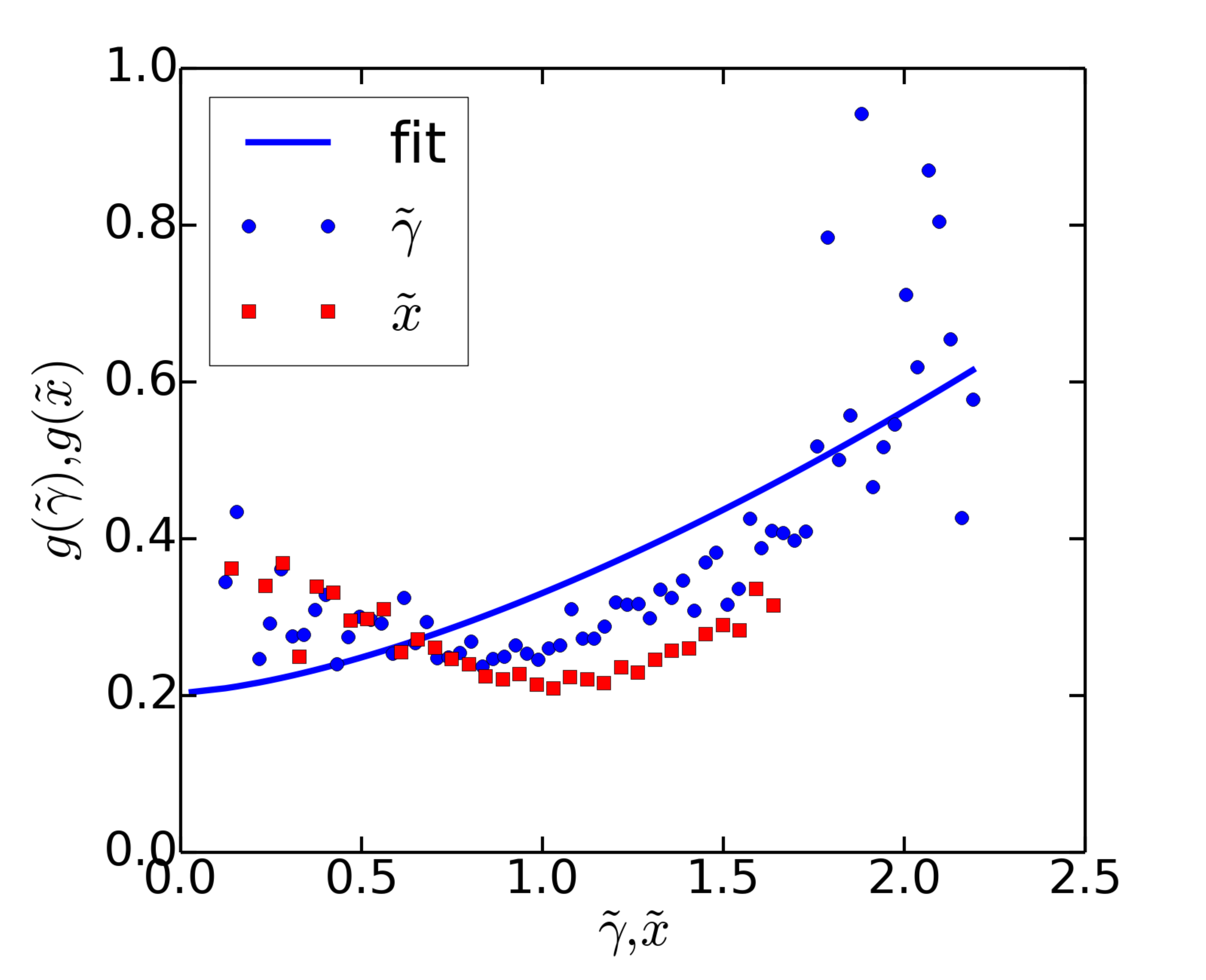}
	\caption{\textbf{Estimated diffusion for the signals $\tilde{\gamma}(t)$ and $\tilde{x}(t)$ for "Starbucks".} Note that the diffusion estimated for the two signals is almost equal for this brand name, therefore making it hard to filter out the effect of the dynamical interest $\gamma(t)$ present in the signal of $\tilde{\gamma}(t)$.}
	\label{fig:starbucks g}
\end{figure}

The variable $\tilde{\gamma}(t)$ is the closest approximation we get of $\gamma(t)$ by our analysis. To see the effect of the Poisson statistics on our analysis, we have also generated the signal

\begin{align*}
\tilde{x}(t) &\equiv \frac{\mathrm{Pois}\left(\left\langle x \right\rangle_D\!(t)\right)}{\left\langle x \right\rangle_D\!(t)}, \\
&= \frac{\mathrm{Pois}\left( \left\langle \gamma \right\rangle_T A(t)\right)}{ \left\langle \gamma \right\rangle_T A(t) }, \numberthis \label{eq:tilde_x}
\end{align*}
and applied the algorithm to both $\tilde{\gamma}(t)$ and $\tilde{x}(t)$. The variable $\tilde{x}(t)$ is equivalent to $\tilde{\gamma}(t)$, but with the dynamics of $\gamma(t)$ replaced by the mean value $\left<\gamma\right>_T$ (compare Eqs. (\ref{eq:tilde_gamma}) and (\ref{eq:tilde_x})). By applying the algorithm to both $\tilde{\gamma}(t)$ and $\tilde{x}(t)$, we hope to be able to separate the effect of the daily variation and the Poisson statistics from the actual dynamics of $\gamma(t)$.

In Fig.~\ref{fig:samsung analysis}A we show $\tilde{\gamma}(t)$ and the corresponding instance of $\tilde{x}(t)$ for ``Samsung''. Note that the Poisson noise of $\tilde{x}(t)$ is not enough to explain the bursty behavior observed for the tweet rates. The resulting drift (Fig.~\ref{fig:samsung analysis}B) and diffusion (Fig.~\ref{fig:samsung analysis}C) terms are estimated using our algorithm on the two signals shown below. The drift of $\tilde{\gamma}(t)$ has been fitted with a function on the form
\begin{equation}\label{eq:fit1}
f\left(\tilde{\gamma}\right) = a_f - b_f\tilde{\gamma}.
\end{equation}
A best fit yields the coefficients $a_f = 0.47$ and $b_f = 0.43$. Similarly, the diffusion has been fitted using a function on the form
\begin{equation}\label{eq:fit2}
g\left(\tilde{\gamma}\right) = a_g + b_g\tilde{\gamma}^{3/2},
\end{equation}
with  $a_g = 0.31$ and $b_g = 0.23$. The analysis has been performed in dimensionless time, $t \rightarrow t / \Delta t$, such that $dt=1$. We have been unable to estimate the error bars in the presence of the Poisson statistics. 

Our algorithm estimates a linear drift term for both the data, $\tilde{\gamma}(t)$, and for the synthetic signal, $\tilde x (t)$. For $\tilde{x}(t)$, we find a coefficient $b_f = 1$, which is expected from a Poisson process. The fact that we find $b_f = 0.43 $ for $\tilde{\gamma}(t)$ shows that the data is more rich than a simple homogeneous or a weakly inhomogeneous Poisson process. In other words, the fluctuations of ${\gamma}(t)$ are comparable or stronger than the fluctuations generated by the superimposed Poisson process. While we cannot quantify the influence of the Poisson process on the linear drift, we are confident that $\tilde{\gamma}$=1 is the only stationary point of $\tilde\gamma$, corresponding to a potential, $V(\gamma) = \int_\gamma f(\gamma') d\gamma'$, with just a single minimum. Furthermore, we do not expect the drift to depart significantly from a linear form around and above the fix point. 
A drift term of this form limits the signal and allows bursts to be generated by the multiplicative diffusion term.

In the plot showing the diffusion terms, we find that the effect of the Poisson statistics is very distinctly visible as a constant background noise. It indeed matches the size of the coefficient $a_g$ pretty well. We therefore propose that the this first term is due Poisson noise and therefore that the underlying variable $\gamma(t)$ is described solely by the second term, $g(\gamma)=b\gamma^{3/2}$.

We apply the same analysis to the other brand names and provide in Table~\ref{tbl:r^2} the length of the fitted data series, $N_\mathrm{data}$, the mean tweet rate, $\left<x\right>$, the ratio between the maximum and minimum of the daily variation, $\mathrm{DV} \equiv \mathrm{max} \left[\left\langle x \right\rangle_D\!(t)\right] / \mathrm{min} \left[\left\langle x \right\rangle_D\!(t)\right]$ , and the goodness-of-fit values for the diffusion term, $R^2$. We find that the fit captures the observed diffusion well for 4 of the 7 brand names, but it performs poorly for the last 3. We believe that this is the result of applying the algorithm to a limited time series under the effects of daily variation and Poisson noise. As an example of this, we show in Fig. \ref{fig:starbucks g} the result of applying the algorithm to the tweet rate signal of "Starbucks". We see that a Poisson process captures most of the fluctuations found in the dynamical signal, i.e.~the diffusion terms of $\tilde{\gamma}$ and $\tilde{x}$ are approximately equal. We therefore conclude that the average interest, $\left<\gamma\right>$, the big daily variation, $A(t)$, and the Poisson noise is enough to explain most of the signal for "Starbucks". This leaves very little room in the analysis to capture the dynamics of $\gamma(t)$ (compare with Fig. \ref{fig:samsung analysis}) and may explain the poor performance of the fit.

\begin{table}
\begin{center}
    \begin{tabular}{| l | l | l | l | l |}
    \hline
    Brand & $N_\mathrm{data}$ & $\left< x \right>$ & DV & $R^2$ \\ \hline
    Samsung & 74,646 & 10.2 & 1.8 & 0.96  \\ \hline
    Google & 81,105 & 67.9 & 1.8 & 0.95  \\ \hline
    Gucci & 74,715 & 19.3 & 8.7 & 0.91  \\ \hline
    BMW & 79,599 & 3.1 & 2.4 & 0.86  \\ \hline
    Heineken & 179,607 & 1.5 & 3.2 & 0.74  \\ \hline
    Starbucks & 72,570 & 24.0 & 5.3 & 0.57  \\ \hline
    Pepsi & 57,548 & 7.2 & 4.8 & 0.53  \\ \hline
    \end{tabular}
\end{center}
\caption{\textbf{Brand name characteristics} For each brand name we show the length of the time series, $N_\mathrm{data}$, the mean tweet rate, $\left< x \right>$, the ratio between the minimum and maximum of the daily variation, "DV", and the $R^2$-values for the fit of Eq. \ref{eq:fit2} to the diffusion terms estimated by the algorithm.}
\label{tbl:r^2}
\end{table}

In general, however, we find that the analysis of $\tilde{\gamma}(t)$ provides support to the hypothesized noise exponent of 3/2. We therefore propose that the global user interest is described by the following model

\begin{equation}\label{eq:model}
d\gamma(t) = f(\gamma)dt + b\gamma^{3/2}dW,
\end{equation}
where $f(\gamma)$ is a slowly decreasing drift term derived from a single well potential. \textcolor{black}{We emphasize that in order to derive this result, we assume that $\gamma(t)$ is described by the stochastic differential equation, Eq.~(\ref{eq:general sde}), and that $A(t)$ can be approximated by a periodic function with a daily period. Finally, our method works best when the Poisson fluctuations are not too strong.}

In the next section we show that if the single well potential defining $f(\gamma)$ is approximated by an infinite well, we obtain a probability distribution with a power law exponent of -3 and a power spectrum with a power law exponent of -1. This is in agreement with the characteristic behavior of the brand name signals analyzed in \cite{mathiesen2013excitable}.

\section*{Distribution and power spectrum from model}
To derive the probability distribution and power spectrum for the model proposed in Eq. (\ref{eq:model}) we switch from the Langevin equation to the corresponding Fokker-Planck formulation

\begin{equation}\label{eq:fokker-planck}
\partial_t P(\gamma,t) = \partial^2_\gamma \left(\frac{b^2\gamma^3}{2}  P(\gamma,t)\right).
\end{equation}
Here we have approximated the drift potential by an infinite well. This  yields a vanishing drift in the region $\gamma \in \left[\gamma_{\mathrm{min}},\gamma_{\mathrm{max}}\right]$ and reflective boundaries at the effective potential walls $\gamma_{\mathrm{min}}$ and $\gamma_{\mathrm{max}}$. One finds the stationary distribution

\begin{equation}
P_s(\gamma) = \frac{N}{\gamma^3}
\end{equation}
where $N$ is the normalization constant. The same asymptotic power law is found in the case of a linear drift, which is promising since it matches the behavior of the data.

Eq. (\ref{eq:fokker-planck}) may be solved using the method of eigenfunctions as explained in \cite{Kaulakys2010}. One finds that for an intermediate range of frequencies the power spectrum scales as
\begin{equation}
S(f) \sim \frac{1}{f},
\end{equation}
which is also the case for the data.

\begin{figure}[htb]
	\centering
	\includegraphics[width=0.5\textwidth]{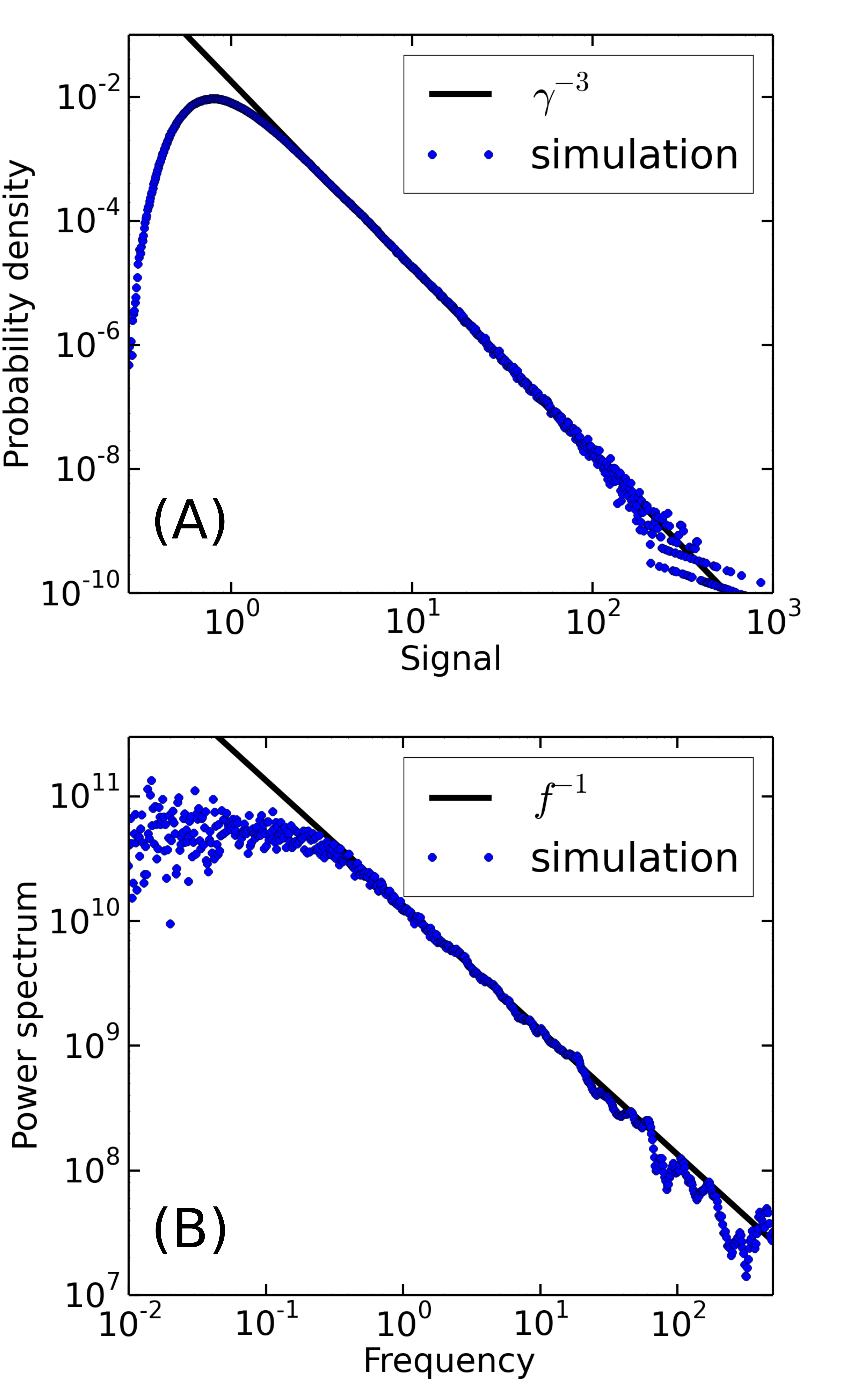}
	\caption{\textbf{Plots of the probability density function (A) and power spectrum (B) for a simulation of the model in Eq. (\ref{eq:simulated model})}. Note that the power law exponents of -1 and -3 match those of the data.}\label{fig:simulationsresults}
\end{figure}

The model proposed for the dynamics of interest, Eq.~(\ref{eq:model}), is therefore successful at simultaneously explaining the scaling exponents of the signal distribution and the corresponding power spectrum. To show the validity of the infinite well approximation, we conclude the paper with a simulation of the model with a linear drift

\begin{equation}\label{eq:simulated model}
d\gamma = (1-0.1\gamma)dt + \gamma^{3/2}dW,
\end{equation}
An efficient and accurate numerical integration may be performed by considering the inverse variable, $\tau = 1 / \gamma$, which may be integrated using the splitting up method \cite{bensoussan1992approximation}. In Fig. \ref{fig:simulationsresults} we show the distribution and power spectrum for the simulated signal, and we observe that the linear drift is consistent with the power laws observed in the data. The exponent of the distribution is fitted to $\alpha_1 = -2.961\pm0.002$ using the \textcolor{black}{the maximum likelihood routine introduced in \cite{alstott2014powerlaw}. The exponent of the power spectrum is found to be $\alpha_2 = -0.98\pm0.03$ using a logarithmic binning and a least squares fit. The corresponding errorbars are estimated by bootstrapping.}

\section*{Conclusion}
In this paper we have studied the dynamics of interest in global brands by analyzing tweet rates on the online social media site Twitter. As a result of the correlations in the user behavior, the rates are found to be bursty and distributed as a power law with an exponent of -3 and have a power spectrum inversely proportional to the frequency. Since the global interest in a brand name is the result of many random events, we have proposed to model it by a stochastic differential equation with a simple drift and a diffusion like term. By analyzing the fluctuations in the tweet rate signals, we find that the diffusion term scales like a power with an exponent of $3/2$. The derived diffusion term may explain the pronounced burstiness and the 1/f noise observed for the tweet rate signals.

It remains an open question whether the dynamics observed for the brand names on Twitter can also be observed for the occurrence of other keywords or even in other large social organizations?  Another interesting question, which we have not addressed with our model is, what is the detailed behavior of individual humans that leads to correlated behavior given by our model? In general, the growing information available on human behavior in global-scale social organizations has helped answer parts of these questions and further analysis along the lines of this paper might provide a more complete picture.

\bibliographystyle{ieeetr}
\bibliography{reference}

\begin{thebibliography}{10}

\bibitem{king2011ensuring}
G.~King, ``Ensuring the data-rich future of the social sciences,'' {\em
  Science(Washington)}, vol.~331, no.~6018, pp.~719--721, 2011.

\bibitem{osborne2014facebook}
M.~Osborne and M.~Dredze, ``Facebook, twitter and google plus for breaking
  news: Is there a winner?,'' in {\em Proceedings of the International
  Conference on Weblogs and Social Media}, 2014.

\bibitem{hermida2014sourcing}
A.~Hermida, S.~C. Lewis, and R.~Zamith, ``Sourcing the arab spring: A case
  study of andy carvin\'s sources on twitter during the tunisian and egyptian
  revolutions,'' {\em Journal of Computer-Mediated Communication}, vol.~19,
  no.~3, pp.~479--499, 2014.

\bibitem{mathiesen2013excitable}
J.~Mathiesen, L.~Angheluta, P.~T. Ahlgren, and M.~H. Jensen, ``Excitable human
  dynamics driven by extrinsic events in massive communities,'' {\em
  Proceedings of the National Academy of Sciences}, vol.~110, no.~43,
  pp.~17259--17262, 2013.

\bibitem{alstott2014powerlaw}
J.~Alstott, E.~Bullmore, and D.~Plenz, ``powerlaw: a python package for
  analysis of heavy-tailed distributions,'' {\em PloS one}, vol.~9, no.~1,
  p.~e85777, 2014.

\bibitem{garas2012emotional}
A.~Garas, D.~Garcia, M.~Skowron, and F.~Schweitzer, ``Emotional persistence in
  online chatting communities,'' {\em Scientific Reports}, vol.~2, 2012.

\bibitem{kwak2010twitter}
H.~Kwak, C.~Lee, H.~Park, and S.~Moon, ``What is twitter, a social network or a
  news media?,'' in {\em Proceedings of the 19th international conference on
  World wide web}, pp.~591--600, ACM, 2010.

\bibitem{myers2014information}
S.~A. Myers, A.~Sharma, P.~Gupta, and J.~Lin, ``Information network or social
  network?: the structure of the twitter follow graph,'' in {\em Proceedings of
  the companion publication of the 23rd international conference on World wide
  web companion}, pp.~493--498, International World Wide Web Conferences
  Steering Committee, 2014.

\bibitem{Sakaki:2010:EST:1772690.1772777}
T.~Sakaki, M.~Okazaki, and Y.~Matsuo, ``Earthquake shakes twitter users:
  Real-time event detection by social sensors,'' in {\em Proceedings of the
  19th International Conference on World Wide Web}, WWW '10, (New York, NY,
  USA), pp.~851--860, ACM, 2010.

\bibitem{DBLP:journals/corr/abs-1010-3003}
J.~Bollen, H.~Mao, and X.-J. Zeng, ``Twitter mood predicts the stock market,''
  {\em CoRR}, vol.~abs/1010.3003, 2010.

\bibitem{DBLP:journals/corr/abs-1003-5699}
S.~Asur and B.~A. Huberman, ``Predicting the future with social media,'' {\em
  CoRR}, vol.~abs/1003.5699, 2010.

\bibitem{kobayashi19821}
M.~Kobayashi and T.~Musha, ``1/f fluctuation of heartbeat period,'' {\em
  Biomedical Engineering, IEEE Transactions on}, no.~6, pp.~456--457, 1982.

\bibitem{voss1992evolution}
R.~F. Voss, ``Evolution of long-range fractal correlations and 1/f noise in dna
  base sequences,'' {\em Physical review letters}, vol.~68, no.~25, p.~3805,
  1992.

\bibitem{weissman19881}
M.~Weissman, ``1/f noise and other slow, nonexponential kinetics in condensed
  matter,'' {\em Reviews of modern physics}, vol.~60, no.~2, p.~537, 1988.

\bibitem{keshner19821}
M.~S. Keshner, ``1/f noise,'' {\em Proceedings of the IEEE}, vol.~70, no.~3,
  pp.~212--218, 1982.

\bibitem{Barabasi:2005}
A.-L. Barabasi, ``The origin of bursts and heavy tails in human dynamics,''
  {\em Nature}, vol.~435, p.~207, 2005.

\bibitem{horvath2014spreading}
D.~X. Horv{\'a}th and J.~Kert{\'e}sz, ``Spreading dynamics on networks: the
  role of burstiness, topology and non-stationarity,'' {\em New Journal of
  Physics}, vol.~16, no.~7, p.~073037, 2014.

\bibitem{siegert1998analysis}
S.~Siegert, R.~Friedrich, and J.~Peinke, ``Analysis of data sets of stochastic
  systems,'' {\em Physics Letters A}, vol.~243, no.~5, pp.~275--280, 1998.

\bibitem{gardiner2009stochastic}
C.~Gardiner, {\em Stochastic Methods: A Handbook for the Natural and Social
  Sciences}.
\newblock Springer Series in Synergetics, Springer, 2009.

\bibitem{Kaulakys2010}
J.~Ruseckas and B.~Kaulakys, ``1/f noise from nonlinear stochastic differential
  equations,'' {\em Phys. Rev. E}, vol.~81, p.~031105, Mar 2010.

\bibitem{bensoussan1992approximation}
A.~Bensoussan, R.~Glowinski, and A.~Ra{\c{s}}canu, ``Approximation of some
  stochastic differential equations by the splitting up method,'' {\em Applied
  Mathematics and Optimization}, vol.~25, no.~1, pp.~81--106, 1992.

\end{thebibliography}

\pagebreak
\noindent
Supporting Information S1 (pdf file). {\bf Information on the data collection.} 

\vspace{.6cm}
\noindent
Supporting Information S2 (Compressed gzip file of a tar archive). {\bf Timestamp data for tweets.}

\end{document}